\documentclass[10pt,twocolumn]{article}

\usepackage[utf8]{inputenc}
\usepackage[T1]{fontenc}

\usepackage{graphicx}
\usepackage{pdfpages}
\usepackage{comment}

\usepackage{lipsum}
\usepackage{xcolor}

\usepackage{authblk}
\usepackage[margin=2cm]{geometry}
\usepackage{mathtools, cuted}

\usepackage{bm}
\usepackage{amsfonts}
\usepackage{amsmath}
\usepackage{amssymb}
\usepackage{mathrsfs} % cursive letters
\usepackage{siunitx}

\usepackage{mdframed}
\usepackage{nomencl}
\usepackage{multicol}

\makenomenclature

\title{Measuring relative humidity from evaporation with a wet-bulb thermometer: the psychrometer}
\author[1]{Marie Corpart}
\author[1]{Frédéric Restagno}
\author[1]{François Boulogne}
\affil[1]{Université Paris-Saclay, CNRS, Laboratoire de Physique des Solides, 91405, Orsay, France.}

\date{\today}
\begin{document}
\nomenclature{\({\cal R}_{\rm H}\)}{Relative humidity, (Eq.~\ref{eq:def_RH})}
\nomenclature{\(p\)}{Partial pressure of water in ambient air, Pa}
\nomenclature{\(p_{\rm sat}\)}{Saturating pressure of water, Pa, (Eq.~\ref{eq:Antoine_equation})}
\nomenclature{\(R\)}{Radius of spherical bulbs, m}
\nomenclature{\(T_{\rm dry}\)}{Dry bulb temperature, K}
\nomenclature{\(T_{\rm wet}\)}{Wet bulb temperature, K}
\nomenclature{\(Q_{\rm ev}\)}{Heat flux due to evaporation $\Delta_\text{vap}H \, \Phi_{\rm ev}$, W}
\nomenclature{\(Q_{\rm h}\)}{Convective heat flux, W, (Eq.~\ref{eq:Q_h})}
\nomenclature{\(Q_{\rm rad}\)}{Radiative heat flux, W, (Eq.~\ref{eq:Q_rad})}
\nomenclature{\(U\)}{Airflow velocity, ${\rm m} \cdot {\rm s}^{-1}$}
\nomenclature{\(P\)}{Atmospheric pressure, Pa}
\nomenclature{\(\Delta p^\star\)}{Partial pressure difference $p - p_{\rm sat}(T_{\rm wet})$, Pa}
\nomenclature{\(\Delta c^\star\)}{Vapor concentration difference $c_\infty - c_{\rm sat}(T_{\rm wet})$, ${\rm mol} \cdot {\rm m}^{-3}$}

\nomenclature{\(\Delta T^\star\)}{Temperature difference $T_{\rm dry} - T_{\rm wet}$, K}
\nomenclature{\(\Delta_\text{vap}H\)}{Enthalpy of vaporization, ${\rm J} \cdot {\rm mol}^{-1}$}
\nomenclature{\(c_\textrm{sat}\)}{Saturating vapor concentration, ${\rm mol} \cdot {\rm m}^{-3}$}
\nomenclature{\(c_\infty\)}{Vapor concentration in ambient air, ${\rm mol} \cdot {\rm m}^{-3}$}
\nomenclature{\({\cal A}\)}{Psychrometric coefficient, K$^{-1}$, (Eq.~\ref{eq:psychrometer_coeff_model})}
\nomenclature{\({\cal A}^{\rm lim} \)}{Psychrometric coefficient in the limit $U \to \infty$, K$^{-1}$, (Eq.~\ref{eq:A_lim})}

\nomenclature{\({\rm Re}\)}{Reynolds number}
\nomenclature{\({\rm Sc}\)}{Schmidt number}
\nomenclature{\({\rm Pr}\)}{Prandtl number}

\nomenclature{\(\mathcal{D}_{\rm w}\)}{Diffusion coefficient   of the water vapor in air, ${\rm m}^2 \cdot {\rm s}^{-1}$}
\nomenclature{\(\alpha_{\rm air}\)}{Air thermal diffusivity, ${\rm m}^2 \cdot {\rm s}^{-1}$}
\nomenclature{\(\nu_{\rm air}\)}{Air kinematic viscosity, ${\rm m}^2 \cdot {\rm s}^{-1}$}
\nomenclature{\({\cal D}_{\rm w}\)}{Vapor diffusion coefficient, ${\rm m}^2 \cdot {\rm s}^{-1}$}
\nomenclature{\(\Phi_{\rm ev}\)}{Convective evaporation rate, ${\rm mol} \cdot {\rm s}^{-1}$, (Eq.~\ref{eq:Q_ev})}
\nomenclature{\(f_{\rm ev}\)}{Ventilation coefficient for mass transfer}
\nomenclature{\(\beta_{\rm ev}\)}{Numerical constant in the expression of $f_{\rm ev}$}
\nomenclature{\(f_{\rm h}\)}{Ventilation coefficient for heat transfer}
\nomenclature{\(\beta_{\rm h}\)}{Numerical constant in the expression of $f_{\rm h}$}
\nomenclature{\(\lambda_{\rm air}\)}{Dry air thermal conductivity, ${\rm W} \cdot {\rm m}^{-1} \cdot {\rm K}^{-1}$}
\nomenclature{\(\sigma\)}{Stefan-Boltzmann constant, ${\rm W} \cdot {\rm m}^{-2} \cdot {\rm K}^{-4}$}
\nomenclature{\(\epsilon\)}{Emissivity}
\nomenclature{\({\cal R}\)}{Ideal gas constant,  ${\rm J} \cdot {\rm mol}^{-1} \cdot {\rm K}^{-1}$}
\nomenclature{\(\rho_{\rm air}\)}{Mass density of dry air,  ${\rm kg} \cdot{\rm m}^{-3}$}
\nomenclature{\(M_{\rm air}\)}{Molar mass of dry air,  ${\rm kg} \cdot {\rm mol}^{-1}$}
\nomenclature{\(T_{\rm c}\)}{Characteristic temperature, K, (Eq.~\ref{eq:T_c})}
\nomenclature{\(w\)}{Specific humidity, (Eq.~\ref{eq:specific_humidity})}

\twocolumn[
    \begin{@twocolumnfalse}
        \maketitle
        \begin{abstract}
        Measuring the relative humidity of air is an important challenge for meteorological measurements, food conservation, building design, and evaporation control, among other
        applications. Relative humidity can be measured with a
        psychrometer, which is a hygrometer composed of two identical thermometers.
        The bulb of one thermometer is covered by a wick soaked with water so that evaporative cooling  makes it indicate a lower temperature than the dry-bulb thermometer;
       it is possible to determine the relative humidity from the difference between these readings. We describe both a model and an experimental setup to illustrate the principle of a psychrometer for a pedagogical laboratory.
        The science of psychrometry could be more broadly taught at the undergraduate level to help introduce students to aspects of
        measurement techniques, fluid mechanics, heat transfer, and non-equilibrium thermodynamics.
        \end{abstract}
    \end{@twocolumnfalse}
]

\section{Introduction}

Humidity is a general concept referring to the water content of a gas as defined, for example, on page 302 of Ref.  \cite{WMO1966}.
In practice, the relative humidity $\mathcal{R}_{H}$ is the ratio, often expressed as a percentage, of the partial pressure of water in the atmosphere $p$ at temperature $T$ to the saturation vapor pressure $p_\text{sat}(T)$ at the same temperature:
\begin{equation}\label{eq:def_RH}
    \mathcal{R}_{H}(T)=\frac{p(T)}{p_\text{sat}(T)}.
\end{equation}

The measurement and control of relative humidity is of importance in a diversity of fields, including
weather forecasting, health care, building design, conservation, and food processing and preservation. \cite{Norbaeck2014, Camuffo2014, international_commission_on_microbiological_specifications_for_foods_microbial_2005}

Historically, the measurement of relative humidity has been challenging. A broad range of techniques have been developed, varying from the hair hygrometer invented by Saussure in 1780 to   modern electronic sensors based on
impedance measurements due to absorption in thin films as described in part II of Ref. \cite{Kaempfer2012}.
These modern hygrometers need calibration standards, and  the most fundamental standard used by national calibration laboratories is the gravimetric hygrometer as explained page 185 of Ref. \cite{world2018guide}.
Using this method, a certain amount of dry gas is weighed and compared with the weight of the test gas in the same volume.
From this, the amount of water is determined and vapor pressure calculated.
This method can provide accurate measurements, but such systems are cumbersome, expensive, and impractical for student use.
In view of these limitations, it is common  to use alternative standards to calibrate commercial hygrometers.
According to the World Meteorological Organization (WMO), a fundamental such standard is the psychrometer, as described on page 185 of Ref. \cite{world2018guide}.

A psychrometer consists of two thermometers placed side-to-side.
The surface of the sensing element, the wet bulb,  is covered with a soaked muslin to maintain a thin film of water.
The sensing element of the second thermometer, the dry bulb, is simply exposed to the air.
The principle of the psychrometer was discovered by James Hutton in 1792, and soon thereafter the significance of
the role of air flow in its operation was recognized and quantified \cite{Playfair1997,Ivory1822,Belli1830,Espy1834a,Regnault1853}.
However, mathematically modeling the principle of the psychrometer remained a challenge until the beginning of the 20th century as it
requires understanding of  evaporation, boundary-layer theory, and radiative heat flux.

Recently, Caporalini {\em et al.} described the use of a home-made psychrometer in atmospheric physics courses \cite{Caporaloni2004}.
Their approach was configured to ensure that their psychrometer is in strict accordance with official WMO recommendations.
In this article we describe an approach more oriented  to physics students.
Our intent is twofold: (1) to propose a simple model for predicting the relative humidity from temperature measurements,
with particular emphasis on elucidating the interplay between radiative and convective heat fluxes and the air velocity; and (2) to
describe an affordable apparatus to illustrate the key points of the model and demonstrate how it can be used to measure the relative humidity.
Our model is described in Sect. \ref{sec:model} and the apparatus and measurements in Sect. \ref{sec:measurements}.
Section \ref{Conclusion} offers a few summary remarks.

%%%%%%%%%%%%%%%%%%%%%%%%%%%%%%
%
% Analysis
%
%%%%%%%%%%%%%%%%%%%%%%%%%%%%%%
\section{Model}\label{sec:model}

\subsection{Problem description}

\begin{figure}
    \centering
    \includegraphics[width=\linewidth]{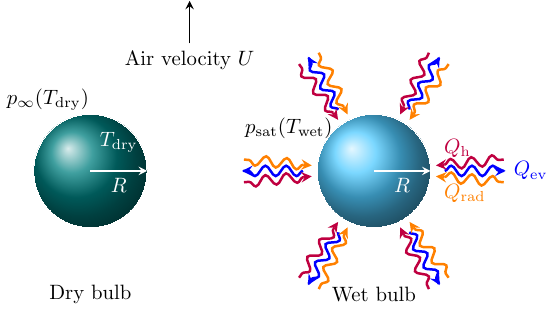}
    \caption{Schematic of the dry and wet bulbs in a steady state regime.
    Bulbs are spherical with radius $R$.
    The dry bulb is the environmental temperature $T_{\rm dry}$ and the evaporating wet bulb is at $T_{\rm wet}<T_{\rm dry}$.
    Evaporating is driven by the vapor pressure difference $p_{\rm sat}(T_{\rm wet}) - p(T_{\rm dry})$.
    The heat flux $Q_{\rm ev}$ due to evaporation compensates the heat fluxes $Q_{\rm h}$ and $Q_{\rm rad}$ by conduction and radiation.
    The system is placed in an air flow characterized by its velocity $U$.}
    \label{fig:model_schematic}
\end{figure}

The analysis in the section will involve a number of quantities which may be unfamiliar to many readers; these are summarized in Table 1.

We model the psychrometer as two spheres of radius $R$ representing the two bulbs placed in an air flow of velocity $U$ as depicted in Fig. ~\ref{fig:model_schematic}.
The atmosphere is characterized by its temperature $T_{\rm dry}$ measured with a dry bulb thermometer, and by its total pressure $P$.
The wet bulb has a temperature $T_\text{wet}$.
We denote $\Delta p^\star = p - p_{\rm sat}(T_\text{wet})$ as the difference of partial vapor pressure between the atmosphere $p(T_{\rm dry})$ and the saturated vapor pressure at the surface of the wet bulb $p_{\rm sat}(T_\text{wet})$.
In practice, the saturated vapor pressure can be calculated with the phenomenological Antoine’s equation
\begin{equation} \label{eq:Antoine_equation}
    p_{\rm sat}(T) = p^\circ\, 10^{A - \frac{B}{C + T}},
\end{equation}
where $p^\circ = 10^5$~Pa and $A$, $B$, $C$ are constants.
This can be used to calculate both $p(T_{\rm dry})$ and $p(T_{\rm wet})$.
For water at $T \in [0, 30]~^\circ$C, $A, B, C$ are obtained by fitting  data extracted from~\cite{Lide2008}, which gives $A = 5.341 \pm 0.003$~K, $B = 1807.5 \pm 1.6$~K, and $C = -33.9 \pm 0.1$~K.

The difference of temperatures $\Delta T^\star=T_\text{dry}-T_\text{wet}$ originates from the evaporation of water on the wet bulb.
Since the molar enthalpy of vaporization $\Delta_\text{vap}H$ is  positive, evaporation cools the wet bulb, so $T_{\rm wet} \leq T_{\rm dry}$.
As a consequence, the vapor concentration at the interface is $p_{\rm sat}(T_{\rm wet})\leq p_{\rm sat}(T_{\rm dry})$.
More precisely, the temperature $T_{\rm wet}$ is the result of the balance between the enthalpy of vaporization and thermal exchanges with the environment.

The psychrometric equation relates the difference of vapor pressures $\Delta p^\star$ and the difference of temperatures $\Delta T^\star$. This has the form
\begin{equation}\label{eq:psychrometric_equation}
  \Delta p^\star  =   - {\cal A} P \Delta T^\star,
\end{equation}
where ${\cal A}$ is the psychrometer coefficient and $P$ is the atmospheric pressure.
Historically, this equation was introduced in the pioneering works of Ivory, August, and Apjohn from considerations of gas expansion, but is now used phenomenologically. ~\cite{Ivory1822,August1825,Apjohn1831,Ferrel1886}.

The purpose of this model is to derive the psychrometric equation (\ref{eq:psychrometric_equation}) with modern concepts, and also to describe the physical origin of the coefficient ${\cal A}$.
The psychometric equation allows a direct calculation of the relative humidity  $\mathcal{R}_{H}(T_\text{dry})$ based on the definition given by equation \ref{eq:def_RH}:
\begin{equation}\label{eq:psychrometric_equationRH}
\mathcal{R}_{H}(T_\text{dry})=\frac{p_{\rm sat}(T_{\rm wet})-{\cal A} P \Delta T^\star}{p_{\rm sat}(T_\text{dry})}.
\end{equation}

\subsection{Mass and heat transfers}

The air flow around the psychrometer is characterized by   the dimensionless Reynolds number $\textrm{Re} = 2UR/\nu_{\rm air}$, where $U$ is the air velocity and $\nu_{\rm air}$ is the kinematic viscosity of the air, which is the ratio between the dynamics viscosity and the fluid density; this is also known as the momentum diffusivity.

The transport of water vapor is driven by the difference of vapor concentrations between the environment and the surface of the wet bulb, \textit{i.e.} $\Delta c^\star = c_\infty - c_{\rm sat}(T_{\rm wet}) < 0$.
This transport is established by  diffusion across the hydrodynamic boundary layer  and is characterized by the diffusion coefficient $\mathcal{D}_{\rm w}$ of the water vapor in air, which has the value  $2.4 \times 10^{-5}$ m$^2$/s at 20~$^\circ$C \cite{Lide2008}.
We also define the Schmidt number as $ \textrm{Sc} = \nu_{\rm air}/ \mathcal{D}_{\rm w} \simeq 0.6$.
The dimensionless Schmidt number is the ratio of momentum diffusivity to mass diffusivity, and is a measure of the relative thickness of the hydrodynamic and mass-transfer boundary layers.
The evaporation rate is
 \begin{equation}\label{eq:Q_ev}
    \Phi_{\rm ev} =  4 \pi R \mathcal{D}_{\rm w} \Delta c^\star f_{\rm ev}.
\end{equation}
The so-called  ventilation coefficient $f_{\rm ev}$ accounts for the effect of the air flow.
There is no exact expression for this coefficient, but Frossling as well as Ranz and Marshall proposed a semi-empirical one:
\begin{equation}\label{eq_fev}
  f_{\rm ev}=1 + \beta_{\rm ev} \textrm{Re}^{1/2} \textrm{Sc}^{1/3},
\end{equation}
where $\beta_{\rm ev}\simeq 0.3$ is a numerical prefactor \cite{Frossling1938,Ranz1952,Ranz1952a}. This expression is suitable for describing experimental results for Reynolds numbers up to 1,280.
The interested reader can find a more general expression in Whitaker \cite{Whitaker1972} which holds for Reynolds numbers up to ${\rm Re} = 7~000$.
The dependence of the ventilation coefficient with the Reynolds number is reminiscent of the boundary layer thickness that scales as ${\rm Re}^{-1/2}$ \cite{Kays1993}.

With Eq. (\ref{eq:Q_ev}), the heat flux due to evaporation is
\begin{equation}
Q_{\rm ev} =  \Delta_\text{vap}H \, \Phi_{\rm ev},\label{eq:Q-phi}
\end{equation}
where $ \Delta_\text{vap}H$ is the enthalpy of vaporization.

Due to the temperature difference between the wet bulb and the environment, heat transfers take place.
Two contributions can be identified.
First, the transfer due to air flow is analogous to the mass transfer formerly described, since  it occurs across a thermal boundary layer.
This heat flux can be expressed as
\begin{equation}\label{eq:Q_h}
    Q_{\rm h} =  4 \pi R \lambda_{\rm air} \Delta T ^\star f_{\rm h},
\end{equation}\label{]eq_fh}
where $\lambda_{\rm air}$ is the air thermal conductivity and $f_{\rm h}$ is the ventilation coefficient \cite{Ranz1952,Ranz1952a}:
\begin{equation}
    f_{\rm h}=1 + \beta_{\rm h} \textrm{Re}^{1/2} \textrm{Pr}^{1/3}.
\end{equation}
The coefficient $\beta_{\rm h}=0.3$ is a numerical prefactor, and the Prandtl number is defined as $ \textrm{Pr} =  \nu_{\rm air} / \alpha_{\rm air} \simeq 0.7$, with $ \alpha_{\rm air} $ being the thermal diffusivity of air.\cite{Ranz1952,Ranz1952a} The Prandtl number is analogous to the Schmidt number for heat transfer.

Second, temperature differences with the environment leads to an energy transfer  $Q_{\rm rad}$ (in W) from radiation given by  Stefan's law,
\begin{equation}\label{eq:Q_rad}
    Q_{\rm rad} = 4\pi R^2 \epsilon \sigma (T_\text{dry}^4 - T_{\rm wet}^4),
\end{equation}
where $\sigma \simeq 5.67 \times 10^{-8}\, {\rm W} \cdot {\rm m}^{-2} \cdot {\rm K}^{-4}$ is the Stefan-Boltzmann constant and $\epsilon$ is the emissivity.
The emissivity $\epsilon$ is close to unity for common materials; for water $\epsilon = 0.96$.\cite{Brewster1992}

As a result, the energy balance
$ Q_{\rm ev} +Q_{\rm h} + Q_{\rm rad} = 0 $ can be written as
\begin{equation}\label{eq:balance}
    Q_{\rm ev} =  - Q_{\rm h} \left( 1 + \frac{Q_{\rm rad}}{Q_{\rm h}} \right),
\end{equation}
which highlights the significance of the radiative flux compared to the conductive one.

From a historical perspective, it is worth noting that Maxwell wrote equations (\ref{eq:Q_ev}) and (\ref{eq:Q_h}) in his description of the psychrometer.\cite{Maxwell1877a}
He also included the effect of radiation, but since the Stefan-Boltzmann was not yet known he assumed radiative flux to simply be linear with the difference of temperatures.\cite{Stefan1879, Boltzmann1884}
This led to a qualitative description of the psychrometer but not a quantitative analysis.

\subsection{Psychrometer coefficient}
To obtain the form of the psychrometric equation (\ref{eq:psychrometric_equation}), two additional steps are necessary.

First, the difference of vapor concentrations $\Delta c^\star$ in Eq. (\ref{eq:Q_ev}) must be expressed as a function of $\Delta p^\star$.
Assuming that vapor is an ideal gas, the difference of vapor molar concentrations can be related to the difference of vapor pressure as
\begin{equation}
    \Delta c^\star = \left(\frac{p}{T_\text{dry}} - \frac{p_{\rm sat}(T_{\rm wet})}{T_{\rm wet}} \right) \frac{1}{{\cal R}},
\end{equation}
 with the molar gas constant ${\cal R} \simeq 8.314$ J$\cdot$mol$^{-1}\cdot$K$^{-1}$.
This can be approximated by $\Delta c^\star \approx \Delta p^\star/( {\cal R} T_{\rm dry})$ for $\Delta T^\star / T_{\rm dry} \ll 1$.
Then we can use the ideal gas law to obtain $\Delta c^\star$ as
\begin{equation}
\Delta c^\star = \Delta p^\star \rho_{\rm air}/P M_{\rm air}, \label{eq:delta_c}
\end{equation}
where $M_{\rm air}$ and $\rho_{\rm air}$ are respectively the molar mass  and the mass density of dry air.

Second, the ratio of radiative and convective heat fluxes
\begin{equation}\label{eq:Qrad_Qconv}
\frac{Q_{\rm rad}}{Q_{\rm h}} = \frac{\epsilon \sigma}{\lambda_{\rm air} f_{\rm h}}  \frac{T_\text{dry}^4 - T_{\rm wet}^4}{T_\text{dry} - T_{\rm wet}} R
\end{equation}
can be simplified for $|T_{\rm wet} - T_\text{dry}| / T_\text{dry}\ll 1$ to the form
\begin{equation}\label{eq:radiation_heat_convection_ratio}
    \frac{Q_{\rm rad}}{Q_{\rm h}}  \simeq  \frac{ T_\text{dry}^3 }{T_{\rm c}^3},
\end{equation}
where we have introduced the characteristic temperature
\begin{equation}\label{eq:T_c}
    T_{\rm c}(R, U) = \left( \frac{{\lambda_{\rm air} f_{\rm h} } }{4R \epsilon \sigma}\right)^{1/3},
\end{equation}
that depends on both the wet bulb size $R$ and the air velocity $U$ through the ventilation coefficient $f_{\rm h}$.

From equation (\ref{eq:balance}), we can obtain the psychrometric equation (\ref{eq:psychrometric_equation}) where we used
equations (\ref{eq:Q_ev}), (\ref{eq:Q-phi}),  and (\ref{eq:delta_c}) to express $Q_{\rm ev}$ as function of $\Delta p^\star$; equation (\ref{eq:Q_h}) for the heat flux proportional to $\Delta T^\star$; and equation (\ref{eq:radiation_heat_convection_ratio}) for the ratio $Q_{\rm rad} / Q_{\rm h}$.
Then, we identify the psychrometer coefficient:

\begin{equation}\label{eq:psychrometer_coeff_model}
 {\cal A} =  \frac{\lambda_{\rm air} M_{\rm air}}{ \Delta_\text{vap}H \, \mathcal{D}_{\rm w} \rho_{\rm air} }    \frac{ f_{\rm h}   }{f_{\rm ev} } \left( 1 + \frac{ T_\text{dry}^3 }{T_{\rm c}^3} \right).
\end{equation}

The psychrometer coefficient is plotted in Fig. \ref{fig:psychrometric_coeff} as a function of the air velocity for four wet bulb sizes.
For wet bulbs of sizes larger than  millimeter scale, we observe a strong decrease of the coefficient with the air velocity, with all curves converging to the same limit.
The physical origin of this behavior is analyzed in the following paragraph.

\subsection{Effect of bulb size and air velocity}
As shown by Eq. (\ref{eq:psychrometer_coeff_model}),  $ {\cal A}$ depends both on the wet bulb size and the air velocity through $f_\text{h}$ and $f_\text{ev}$
although it is  commonly called the psychrometric ``constant.''
Here we analyze the effects of these two parameters.

The ratio $f_{\rm h} / f_{\rm ev}$ can be easily computed from the proposed expressions of Eq. \ref{eq_fev} and \ref{]eq_fh}.
It appears that this ratio depends only weakly  on both $R$ and $U$.
The underlying reason is that the Schmidt and Prandtl numbers are nearly equal for gases, $\textrm{Sc} \approx \textrm{Pr} \approx 1$,
hence we can expect only minute variations of $f_{\rm h} / f_{\rm ev}$. The basis of this similarity is that
kinetic theory shows that the microscopic mechanism of momentum and thermal diffusion in a gas have the same origin.

Next, we can expect the term
$T_\text{dry}^3 / T_{\rm c}^3$
to be negligible if $T_\text{dry}^3 / T_{\rm c}^3\ll 1$.
In practice, we consider that the contribution is small if the correction is smaller than 10\%, \textit{i.e} $T_\text{dry}^3 / T_{\rm c}^3 < 0.1$.
In the limit of vanishing air velocities ($U\rightarrow 0$), the ratio is proportional to $R$ and is negligible for $R < 0.5$~mm, where we used $\lambda_{\rm air} = 0.026 $ $\rm{W\cdot m^{-1} \cdot K^{-1}}$ and $T_\text{dry}=293$ K.
This size is much smaller than bulbs of thermometers, so that in the absence of air flow the psychrometer coefficient is sensitive to the bulb size.
In practice, ensuring the absence of air flow is difficult because of the presence of natural convection, so this limit is not really applicable.

As for  the effect of the air velocity, we observe that $T_\text{dry}^3 / T_{\rm c}^3$ is a decreasing function of $U$.
Thus, for a given wet bulb size, there exists a typical air velocity above which the ratio is negligible, \textit{i.e.} $T_\text{dry}^3 / T_{\rm c}^3\ll 1$.
For $R=1$~cm we find a typical velocity of about $5$~m$\cdot$s$^{-1}$.

\subsection{The importance of psychrometer ventilation}

\begin{figure}
    \centering
    \includegraphics[width=\linewidth]{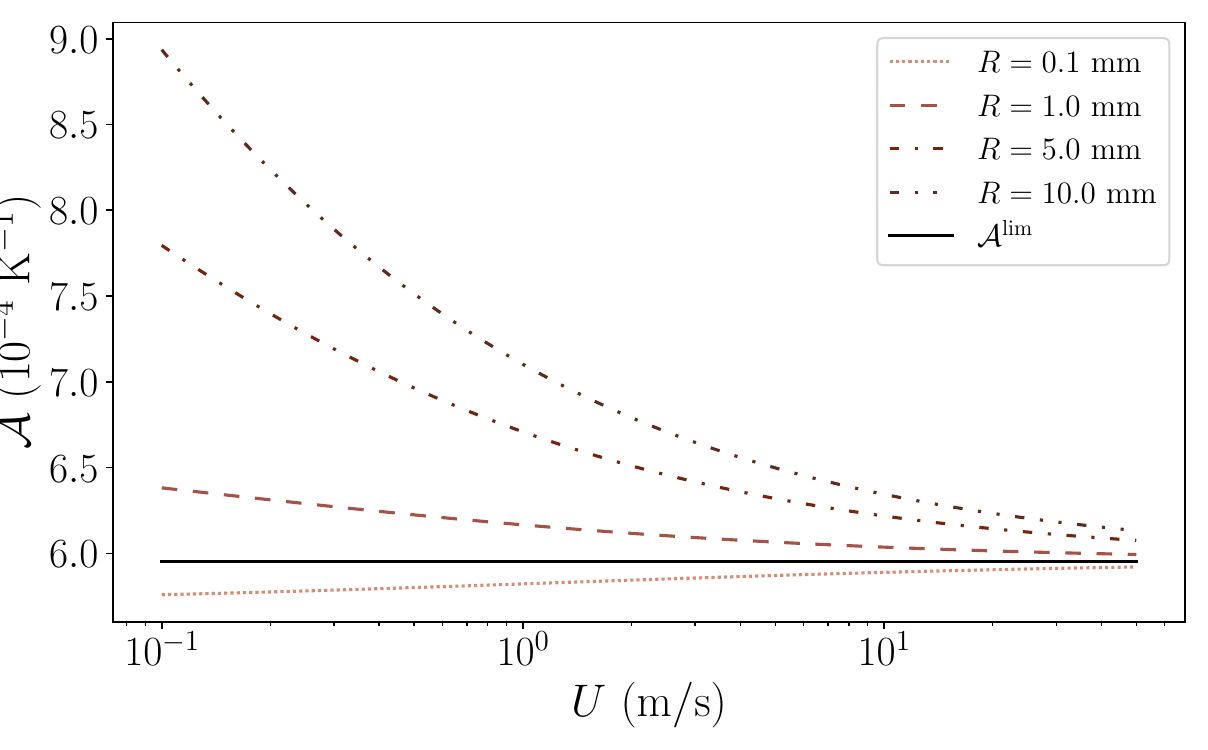}
    \caption{Psychrometer coefficient ${\cal A}$ from equation \ref{eq:psychrometric_equation} as a function of the air speed $U$ and for different radii $R$.
    The black solid line represents $ {\cal A}^{\rm lim} $ obtained from the limit $U\rightarrow \infty$.
    }
    \label{fig:psychrometric_coeff}
\end{figure}

The model developed above highlights how the psychrometer coefficient depends on the geometry of the wet bulb and on the air flow velocity.
The dependency on the wet bulb size arises from a competition between the heat fluxes by radiation and by convection, where the former is proportional to the surface area of the bulb while
the latter is proportional to the characteristic size.
For small objects, the radiative heat transfer is negligible whatever the air velocity.
For increasing air flow velocities, this typical size increases because the heat transfer by convection increases, whereas the radiative flux does not change.
For this reason, ventilating the psychrometer reduces the contribution of radiative heat transfers and makes the psychrometer
coefficient weakly dependent on the bulb size.

Forcing the air flow also minimizes the effect of flows from natural air convection, which causes the variation of the psychrometer
coefficient with air velocity to be particularly important for a centimeter-scale bulb, as illustrated in  Fig.~\ref{fig:psychrometric_coeff}.
In the limit $U \rightarrow \infty$, the effect of radiation becomes negligible, and the psychrometer coefficient reduces to
\begin{equation}\label{eq:A_lim}
 {\cal A}^{\rm lim} =   \frac{\lambda_{\rm air} M_{\rm air}}{ \Delta_\text{vap}H \, \mathcal{D}_{\rm w} \rho_{\rm air}} \left(\frac{ {\rm Pr} }{ {\rm Sc}}\right)^{1/3}.
\end{equation}
In our conditions, ${\cal A}^{\rm lim} = 6.0\times 10^{-4}$~K$^{-1}$.
In textbooks one can find the expression  $ {\cal A} =   \lambda_{\rm air} M_{\rm air} / (\Delta_\text{vap}H \, \mathcal{D}_{\rm w} \rho_{\rm air})$. This is sometimes known as the psychrometric ratio,
and is nearly equal to equation (\ref{eq:A_lim}) since $({\rm Pr/Sc})^{1/3}\simeq 1.04$.

\subsection{Carrier chart}

\begin{figure}
    \centering
    \includegraphics[width=\linewidth]{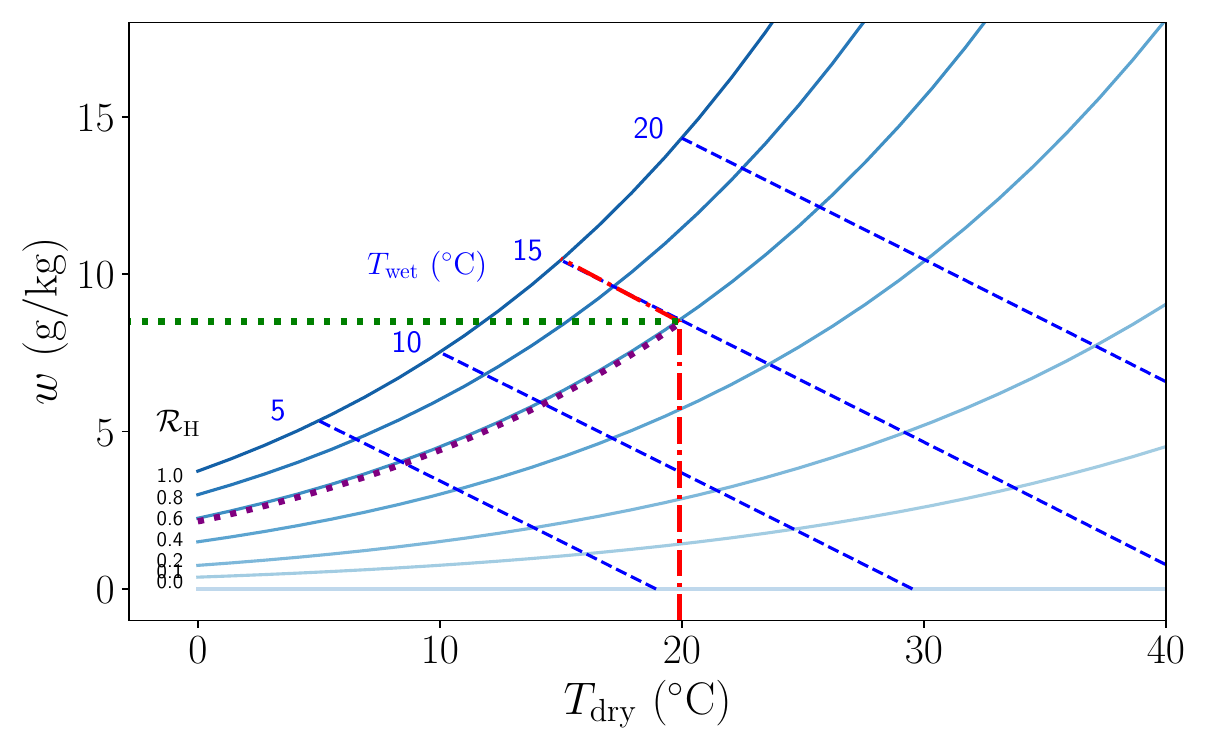}
    \caption{Carrier chart presenting the specific humidity $w$ as a function of the dry bulb temperature in solid lines.
    The dashed lines correspond to the humidity for the specified wet bulb temperature as a function of the dry bulb temperature.
    The chart is produced by using the psychrometer coefficient $ {\cal A}^{\rm lim}$.
    The red lines show the reading for $T_{\rm dry} = 20$~$^\circ$C and $T_{\rm wet} = 15$~$^\circ$C.
    The specific humidity can be read by following the green line and the purple line indicates the relative humidity.}
    \label{fig:Carrier}
\end{figure}

 Now that we have some understanding of the physical origin of the psychrometer coefficient, we are prepared to determine
 the relative humidity from the joint measurement of the dry and wet bulb temperatures.
While one could produce tables which give values of the relative humidity corresponding to pairs of ($T_{\rm dry},T_{\rm wet}$) values,
this would require several pages of output.
Instead, we reproduce a psychrometric chart as originally designed by Carrier, which makes it easy to deduce the relative humidity from the temperatures.\cite{Carrier1911}
This chart also makes it possible to determine other quantities such as the dew point and the absolute humidity, which are relevant in air conditioning.\cite{Gatley2004}
A source code in Python for generating this chart is provided in Supplementary Materials.

Figure~\ref{fig:Carrier} shows a simplified Carrier chart on which we illustrate the following circumstance.
Suppose that a psychrometer indicates a dry bulb temperature of 20~$^\circ$C and a wet bulb temperature of 15~$^\circ$C.
The two red lines originating from each temperature intersect at a point from which the specific humidity (defined in more detail below) can be read from the green dotted line and,
more importantly, the relative humidity from the purple dotted line.
This leads to a relative humidity ${\cal R}_{\rm H} \simeq 60$~\% and a specific humidity $w\simeq 8.5$ g/kg.
For practical use, the chart must be refined by plotting additional curves to obtain a better accuracy.

To construct this chart,
the first step is to plot the water content of air as a function of its temperature $T_{\rm dry}$.
The water content of air depends on temperature since it can vary between zero and the saturating value, which itself depends on the air
temperature; see Eq.  \eqref{eq:Antoine_equation}.
The water content is called specific humidity when defined as the ratio of weight of vapor to the  weight of humid air.
As a good approximation, the specific humidity is close to the mixing ratio, which is the weight of vapor normalized by the weight of dry air.
This approximation originates in the low water vapor pressure compared to the atmospheric pressure, $p_{\rm sat} / P \ll 1$. \cite{Iribarne1981}
From the ideal gas law, the specific humidity can then be written as $w = p M_{\rm w} / (P M_{\rm air})$.
Furthermore, from the definition of the relative humidity (Eq.~\ref{eq:def_RH}) and Antoine's equation (Eq.~\ref{eq:Antoine_equation}), we have
\begin{equation}\label{eq:specific_humidity}
    w(T) = \frac{{\cal R}_{H} M_{\rm w}\,p^\circ\, 10^{A - \frac{B}{C + T}}  }{ P M_{\rm air}},
\end{equation}
which can readily be plotted as a function of the dry bulb temperature for different relative humidity values ${\cal R}_{H}$; this is illustrated in Fig. \ref{fig:Carrier} by the solid lines.

Next, for different given wet bulb temperatures and relative humidity values, we solve the psychrometric equation (Eq. \ref{eq:psychrometric_equationRH}),
for which saturating pressures are calculated from Eq. (\ref{eq:Antoine_equation}) and the psychrometric coefficient is $ {\cal A}^{\rm lim}$ given in Eq. (\ref{eq:A_lim}), ${\cal A}^{\rm lim} = 6.0\times 10^{-4}$~K$^{-1}$ in our conditions.
Equation (\ref{eq:psychrometric_equationRH}) does not directly provide the dry bulb temperature, so we use a Newton-Raphson root-finding algorithm,
 with the wet bulb temperature as a starting estimate.
The optimization returns the dry bulb temperatures, which are represented by the dashed lines in Fig. \ref{fig:Carrier}.
A source code written in Python and using scipy that generates this chart is provided in Supplementary Material.

We emphasize two important points here.
First, the diagram is plotted for a given atmospheric pressure, which is often indicated on charts available in the literature with a ``sea level'' indication.
Second, the chart assumes that radiation effects are negligible.
Thus,  deviations between the predicted wet bulb temperature and the temperature of an evaporating surface may exist, depending on conditions.

%%%%%%%%%%%%%%%%%%%%%%%%%%%%%%
%
% Measurements
%
%%%%%%%%%%%%%%%%%%%%%%%%%%%%%%
\section{Measurements and analysis}\label{sec:measurements}

In this section we test the predictions made in Section~\ref{sec:model} against experimental measurements.
We choose to use a simple apparatus inspired by the design proposed in \cite{Caporaloni2004} to measure the dry and wet
bulb temperatures with a tunable air flow velocity.
This device is affordable for demonstrations and practical student experiments.
A table giving estimated costs of the components and a photograph of the system is provided as Supplementary Material.

\subsection{Experimental setup}

\begin{figure}
    \centering
    \includegraphics[width=.8\linewidth]{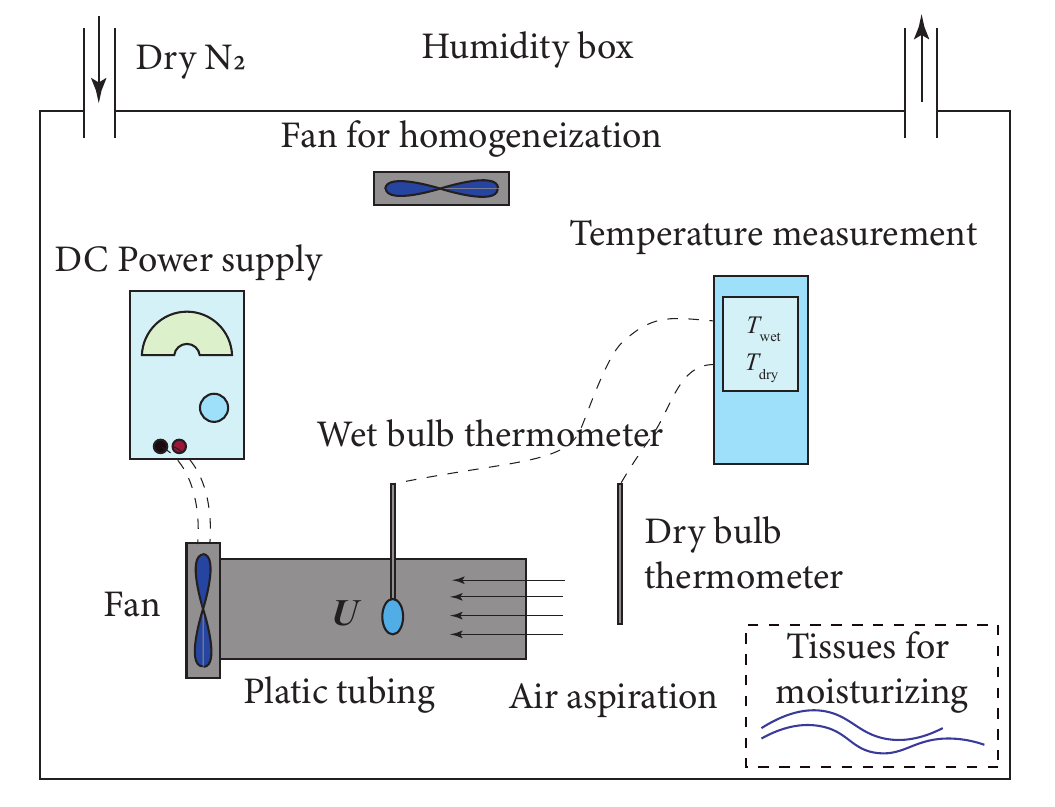}
    \caption{Setup used to measure the temperature difference in air flow produced by an electric fan.
    The tube diameter is about 10~cm.
    }
    \label{fig:fan}
\end{figure}

The psychrometer comprises two Testo type T temperature probes (waterproof, accuracy of $\pm 0.2~^\circ$C),
both connected to a RS PRO 1314 digital interface. One could also use classical alcohol thermometers.
One of the two thermometers is covered with a water-soaked gauze compress.
In our experiments, the wet-bulb has a typical dimension of 8~mm in diameter and 4~cm in length. While our model assumes a spherical bulb,
we  show in the following section that the model gives good results in comparison to experimental data upon
adopting an effective bulb diameter close to the actual one.
We checked at the start and end of the experiment that the gauze is wet.
The first thermometer measures the temperature in the box $T_{\rm dry}$ and the second one measures the wet bulb temperature $T_{\rm wet}$.
As shown in Fig.~\ref{fig:fan}, the two thermometers are placed in a plastic tube equipped with an electric  fan at one end.
Are both within the tube - the dry one is shown outside.
The fan has a diameter of 12~cm and is rated by the manufacturer at 234~${\rm m^3/h}$ (Sunon, EEC0381B1-A99).
The fan is connected to the tube with an adapter made with a 3D-printer to achieve good sealing, although this is not a strict requirement.
By adjusting the voltage applied to the fan, the air flow can be varied, and is measured with a RS PRO AM-4204 hot wire anemometer (maximum measurable velocity  20~m/s; resolution 0.1~m/s).
With this setup we are able to explore air velocities up to 10~m/s, which corresponds to  Reynolds number ${\rm Re} = 2R U / \nu_{\rm air} \simeq 5~000$ for $R=4$~mm.

This device is placed in a transparent glove box measuring $80\times50\times45$~cm) in which the relative humidity can
be adjusted as  described below.
An inlet and an outlet allow flushing the box with dry nitrogen.
The relative humidity in the glove box is measured by using a Velleman DEM501 digital hygrometer with a capacitance sensor (this is not shown in Fig.~\ref{fig:fan}).
We denote $\mathcal{R}_{H}^\text{hygro}$ as the value measured by the hygrometer.
A separate fan placed in the box ensures a rapid homogenization, and experiments are done at ambient temperature.

Starting from ambient humidity, flushing with nitrogen allows drying the air in the box and, with our setup,
allows us to reach humidity values lower than 5~\% in less than 10 minutes.
To increase the humidity we replace the dry nitrogen by ambient air by slightly opening the box.
To reach higher humidities, we put wet towels in the glove box and close the air entrance.
In this way, the relative humidity in the box increases slowly to reach about 95 \%.

\subsection{Measurements}

\begin{figure}[h!]
    \centering
    \includegraphics[width=\linewidth]{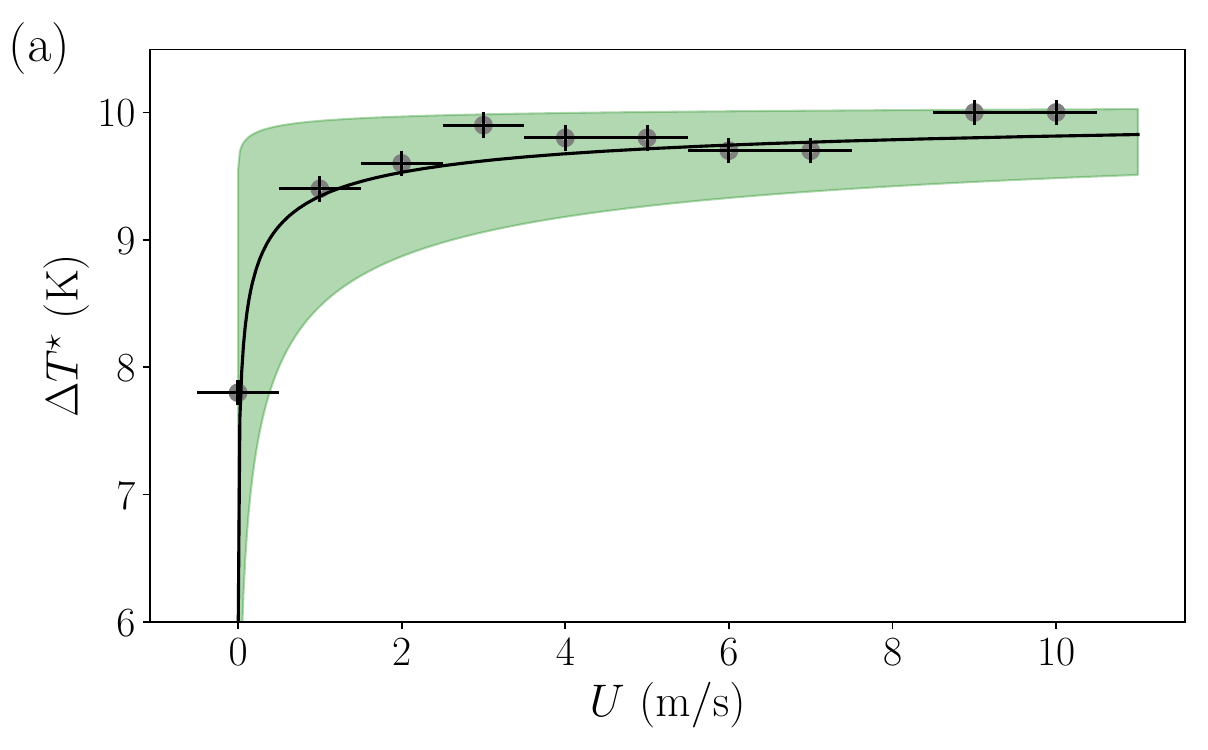}
    \includegraphics[width=\linewidth]{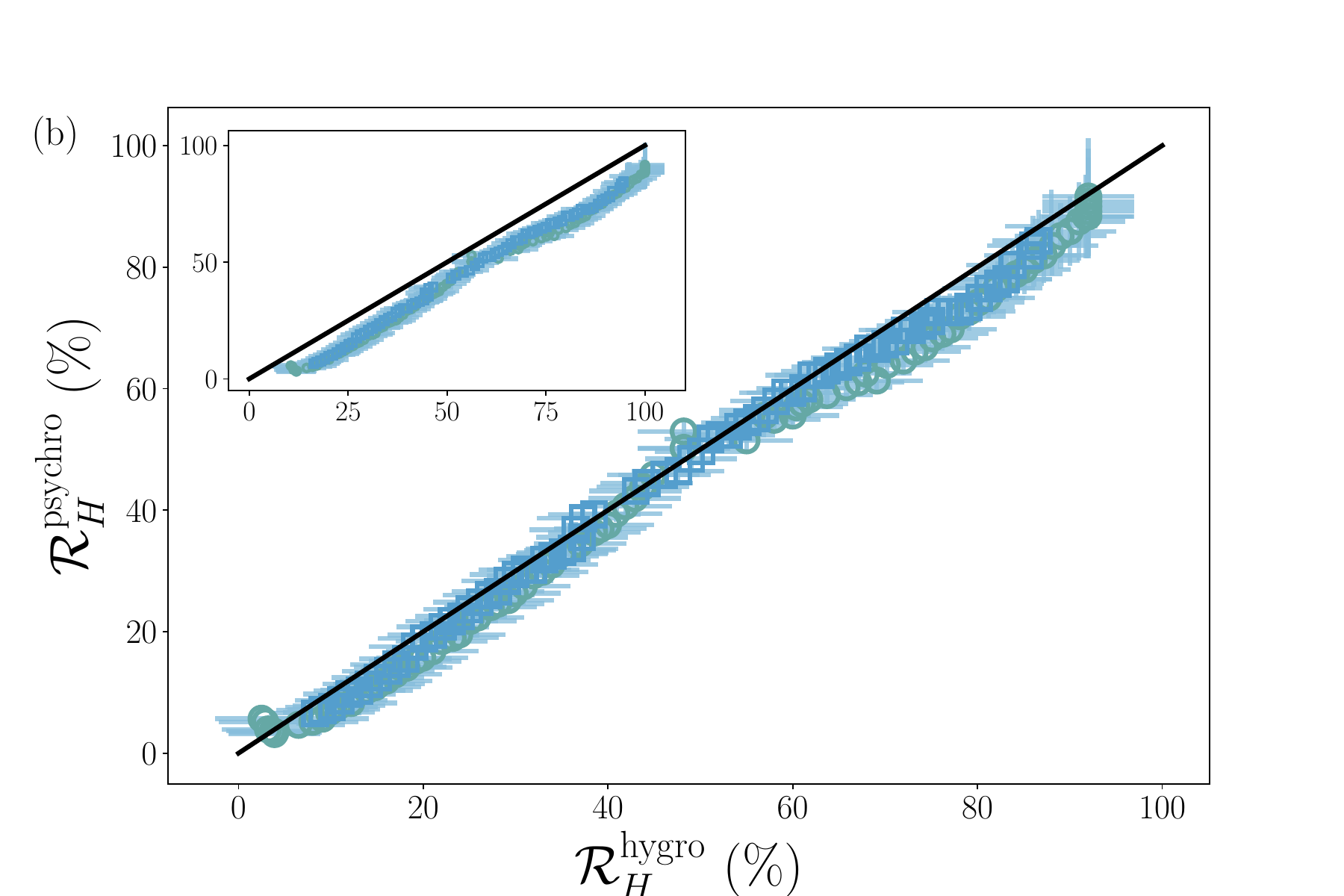}
    \caption{(a) Measured temperature difference $\Delta T^\star$ as a function of the airspeed $U$ for $T_{\rm dry} = 20.3 \pm 0.1$ °C and $\mathcal{R}_{H}^\text{hygro} = 27.8 \pm 2.0$ \% measured with a digital hygrometer.
    The solid curve corresponds to the predicted temperature difference obtained with equations~\ref{eq:psychrometric_equation} and \ref{eq:psychrometer_coeff_model} with a characteristic size $R=1$~cm.
    The green shaded area corresponds to the range $R\in[0.1, 5]$~cm.
    (b) Comparison between the humidity deduced from the psychrometer and the commercial hygrometer.
    The inset shows the data before calibration of the commercial hygrometer and the main plot after.
    The solid black line indicates the equality between axes.
    The main figure is obtained after calibrating the commercial hygrometer with salt solutions as detailed in the text.
    }
    \label{fig:experiments}
\end{figure}

Our first experiment explores the effect of the air flow on temperatures. Here we measured
the temperatures for different applied voltages to the fan once the equilibrium is reached, with the relative humidity held constant.
The dry bulb thermometer proved insensitive to the air flow,
so, we plot in Fig. ~\ref{fig:experiments}(a) the temperature difference $\Delta T^\star$ as a function of the air velocity.
The air flow increases the cooling effect dramatically over the first few meters per second, after which a slower variation is observed.
The solid curve is the prediction of the model of Sect. \ref{sec:model}, which shows a good agreement for a wet bulb size $R=1$~cm.
This size is characteristic of the actual bulb (8 mm diameter), so we did not attempt a more detailed analysis of the effect of the shape of the bulb.
Clearly, for sufficiently large air velocities, the temperature difference becomes independent of the
air flow due to the negligible contribution of the radiative heat flux.
This regime of nearly constant difference thus regime allows a robust measurement of the relative humidity.

To quantify the effect of the humidity on the wet bulb measurement, we  measured the temperature difference
for humidity conditions ranging from $\mathcal{R}_{H}^\text{hygro} = 3$ \% to  $\mathcal{R}_{H}^\text{hygro} = 100$ \% as measured by the commercial hygrometer.
The results are shown in the inset of Fig. ~\ref{fig:experiments}(b).
Measurements are performed from low to high humidity values (green circles) and high to low humidity values (blue squares).
A small hysteresis  of the order of $\delta T = 0.5$~$^\circ$C is observed, which is possibly due to the finite response time of both the thermometer and hygrometer and to the fact that the measurements
are done continuously to avoid a closed-loop setup and ensure a humidity-controlled chamber.

We observe a systematic deviation between the curve  ${\cal R}_{H}^{\rm psychro} = {\cal R}_{H}^{\rm hygro}$ and the experimental data.
Thus, we calibrated the commercial hygrometer by measuring the relative humidity above solutions saturated with various salts,
for which the expected relative humidity is given in \cite{Greenspan1977}.
We performed measurements for KOH solutions for which we expect  ${\cal R}_H = 9$~\%, for MgCl$_2$ (${\cal R}_H = 33$~\%), and for NaCl (${\cal R}_H = 75$~\%).
The values measured by the hygrometer are $\mathcal{R}_{H}^{\rm exp} \approx 15$~\% for KOH, $\mathcal{R}_{H}^{\rm exp} \approx 44$~\% for MgCl$_2$,
and  $\mathcal{R}_{H}^{\rm exp} \approx 83$~\% for NaCl.
The relative humidities measured with the commercial hygrometer are all higher than the expected relative humidity, the difference being about $8$~\%.
Consequently, we calibrated the digital hygrometer by shifting the measurements down by $8$~\%, which yields excellent agreement
with our experimental results as shown in Fig. \ref{fig:experiments}(b).
This procedure indicates that a careful eye must be kept on all measuring equipment:
Digital hygrometers require calibrations and are particularly prone to deviations.
Students should always be prompted to consider calibration methods.

After these measurements and calibrations, it is now possible to propose different teaching situations.
\begin{itemize}
    \item In a course, the teacher can make a single point experiment, show the temperature difference between the dry and wet bulb temperature, and propose the full theoretical calculation (and/or explain how to use a Carrier chart).
    \item In a practical work, the students can perform all the experiments resulting to figure \ref{fig:experiments} simply using a fan in order to be directly in the regime where the air velocity becomes  insignificant.
    \item All the experiments, including the measurement of the effect of the air velocity can be reproduced.
\end{itemize}
The choice between the different pedagogical scenarios whill depend on the level of the student and the purpose of the teaching. The emphasis can be made either on theory or experiments depending of the students level.

%%%%%%%%%%%%%%%%%%%%%%%%%%%%%%
%
% CONCLUSION
%
%%%%%%%%%%%%%%%%%%%%%%%%%%%%%%
\section{Conclusion} \label{Conclusion}

We have described a model for the psychrometer which allows us to quantify the significance of radiative and convective heat transfers
and role of the air velocity on the psychrometer coefficient.
An experimental implementation suitable for student laboratory use gives results in good accord with the model,
as verified by independent measurements with a commercial hygrometer.
Measurements of the relative humidity conducted with air velocities of about a meter per second ensures results reproducible results independent of the air flow.

Quantitative understanding of the relation between evaporation and cooling is an application of practical thermodynamics.
Given concern with the effects of global warming on human health, disasters such as wildfires and hurricanes, and effects on  food and water supplies,
we urge that the concepts of psychrometry be more broadly taught to undergraduates to inform them of how humidity is determined.

\section*{Acknowledgments}
We kindly thank Vincent Klein for creating the 3D-printed adapter, Saint-Gobain and ANRT for funding this study. FR would like to thank Suzanne Lafon for her careful correction of the manuscript.

\section*{Author Declarations}
The authors have no conflicts to disclose.

\twocolumn[
\begin{@twocolumnfalse}
    \begin{mdframed}
        \begin{multicols}{2}
        \printnomenclature
        \end{multicols}
    \end{mdframed}
\end{@twocolumnfalse}
]

\bibliographystyle{ajp}
\bibliography{biblio}

\newpage\clearpage


\section*{Supplementary material}

# Supplementary materials
# Measuring relative humidity from evaporation with a wet-bulb thermometer: the psychrometer

M. Corpart[1], F. Restagno[1], and F. Boulogne[1]

[1]Université Paris-Saclay, CNRS, Laboratoire de Physique des Solides, 91405, Orsay, France.

## 1 Experimental setup

In the following table, we present the material we used for the proposed experiments based from our lab equipment often present in lab for undergraduate students. The price is provided as reference. The total cost can be lowered by selecting a less efficient electronic equipment without compromising measurement quality. The price of affordable alternative equipment are indicated in the table.

| **Equipment** | **Reference** | **Price** |
|---|---|---|
| Fan | Sunon, EEC0381B1-A99 | 26 € |
| Tubing | – | 7 € |
| Power supply | Velleman LABPS6005 | 199 € |
|  | (alt) Eventek 0-30 V 0-5 A CC | 60 € |
| Temperature meter | RS PRO, 1314 | 211 € |
|  | (alt) ThermoPro TP02S | $2 \times 10$ € |
| Temperature probes | Testo, type T | $2 \times 49$ € |
|  | (alt) TP02S has an integrated probe | 0 € |
| Anemometer | Testo 440 | 550 € |
|  | (alt) BTMETER BT-100 | 35 € |
| Hygrometer | Velleman DEM501 | 59 € |
|  | (alt) ICQUANZX 2-en-1 LCD | 7 € |
| Glove box | (alt) Ikea Samla 45 L with cover | 10 € |

The glovebox can be simply made of an ordinary plastic box with small apertures for flushing dry nitrogen and for cables.

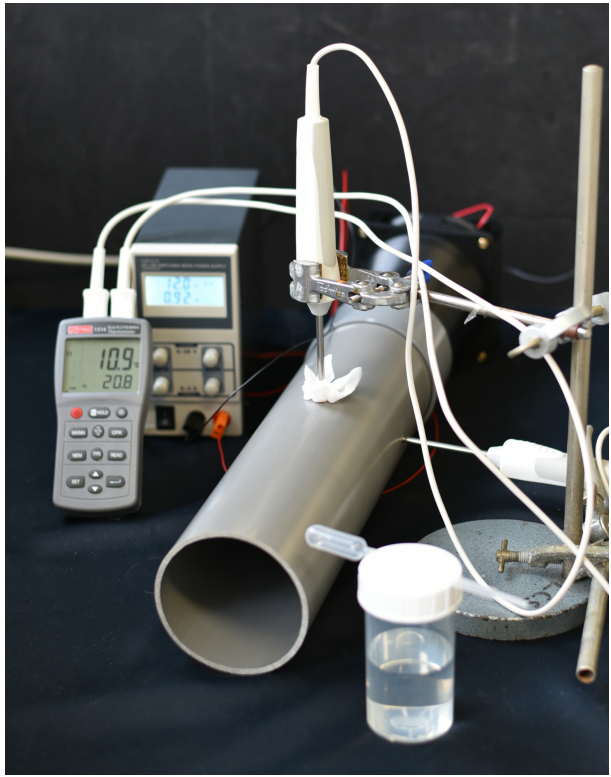

*Photograph of our experimental setup.*

# 2 Source code

[1]:
```python
import numpy as np
from scipy import constants, optimize

from matplotlib import cm
from matplotlib import colors
import matplotlib.pyplot as plt
import matplotlib
%matplotlib inline


def switch_mpl_style(latex=True):
    if latex:
        plt.rcParams['figure.figsize'] = (5 * 1.61, 5)
        matplotlib.rcParams['text.usetex'] = True

        plt.rc('axes', labelsize=22)    # fontsize of the x and y labels
        plt.rc('xtick', labelsize=20)   # fontsize of the tick labels
        plt.rc('ytick', labelsize=20)   # fontsize of the tick labels
        plt.rc('legend', fontsize=15)   # legend fontsize
    else:
        matplotlib.rcParams['text.usetex'] = False
        plt.rc('axes', labelsize=10)    # fontsize of the x and y labels
        plt.rc('xtick', labelsize=10)   # fontsize of the tick labels
        plt.rc('ytick', labelsize=10)   # fontsize of the tick labels
        plt.rc('legend', fontsize=10)   # legend fontsize
```

## 2.1 Constants

[2]:
```python
lambda_air = 0.026 # SI
Delta_H_vap = 2.5 * 1e6 # SI
D = 2.5 * 1e-5 # SI
M_w = 0.018  # SI
M_air = 0.029  # SI
rho_air = 1.2  # SI
P_atm = constants.atmosphere
emissivity = .96
Prandtl = 0.7
Schmidt = 0.6
```

## 2.2 Carrier's chart

### 2.2.1 Pressure and concentrations

```python
[3]: def concentration_ideal_gas(p, T):
         """
         Concentration from pressure and temperature.

         Parameters
         ----------
         p: Pressure in Pa
         T: Temperature in °C
         """
         return p * M_w / (constants.R * constants.convert_temperature(T, 'C', 'K'))


     def p_sat_Antoine(T):
         """
         Vapor pressure from Antoine equation.

         Parameters
         ----------
         T: Temperature in °C.
         """
         T = constants.convert_temperature(T, 'C', 'K')
         p0 = 1e5
         A = 5.3411
         B = 1807.52
         C = -33.901
         return p0 * 10**(A - B / (C + T))


     def relative_humidity(delta_pression, T_i, T_infty):
         return (p_sat_Antoine(T_i) - delta_pression) / p_sat_Antoine(T_infty)
```

### 2.2.2 Psychrometer coeff

```python
[4]: def psychrometric_equation(emissivity, RH, T_wet, T_dry):
         """
         Psychrometric equation to minimize.

         Parameters
         ----------
         emissivity: emissivity
         RH: relative humidity
         T_dry: Temperature of the dry bulb in °C
         T_wet: Temperature of the wet bulb in °C
         """
```

```python
    # Psychrometer coefficient A
    # in the limit U->infty
    A = lambda_air * M_air / (rho_air * M_w * Delta_H_vap * D)
    A *= (Prandtl / Schmidt)**(1/3)

    Delta_T = T_dry - T_wet
    rhs = A * Delta_T * P_atm
    lhs = p_sat_Antoine(T_wet) / p_sat_Antoine(T_dry) - RH
    lhs *= p_sat_Antoine(T_dry)
    return rhs - lhs


def find_T_wet(T_wet, T_dry, RH, emissivity):
    return psychrometric_equation(emissivity, RH, T_wet, T_dry)


def find_T_dry(T_dry, T_wet, RH, emissivity):
    return psychrometric_equation(emissivity, RH, T_wet, T_dry)


def specific_humidity(T_dry, RH, M_w, M_air):
    """
    Compute the mass ratio in kg/kg.

    Parameters
    ----------
    T_dry: Temperature of the dry bulb in °C
    RH: relative humidity
    M_w: molar weight of water
    air_density: Air density (kg/m3)
    """
    vap_pressure = p_sat_Antoine(T_dry) * RH
    return vap_pressure * M_w / M_air / constants.atmosphere
```

### 2.2.3 Chart

```python
[5]: switch_mpl_style(latex=True)

T_dry_space = np.linspace(0, 40, 30)
RH_space = np.array((0, .1, .2, .4, .6, .8, 1))
T_wet_space = np.array((5, 10, 15, 20))
color_map_RH = cm.get_cmap('Blues', len(RH_space) + 5)

# Compute the mass ratios
for i, RH in enumerate(RH_space):
    w = specific_humidity(T_dry_space, RH, M_w, M_air)
    w *= 1e3   # convert from kg/kg to g/kg
```

```python
        plt.plot(T_dry_space, w, color=color_map_RH(i + 3))
        plt.text(-1.7, w[0] * .9, RH)

# Compute ratios for constant wet bulb temperatures
for T_wet in T_wet_space:
    res_spec_hum = np.zeros_like(RH_space)
    res_T_dry = np.zeros_like(RH_space)
    for i, RH in enumerate(RH_space):
        T_dry = optimize.newton(find_T_dry,
                                T_wet,
                                args=(T_wet, RH, emissivity))
        w = specific_humidity(T_dry, RH, M_w, M_air)
        w *= 1e3  # convert from kg/kg to g/kg
        res_spec_hum[i] = w
        res_T_dry[i] = T_dry

    plt.plot(res_T_dry, res_spec_hum, 'b--')
    plt.text(res_T_dry[-1] - 2,
             res_spec_hum[-1],
             T_wet,
             color='blue',
             fontsize=15)

plt.xlabel(r'$T_{\rm dry}$ $(^\circ {\rm C})$')
plt.text(7,
         11.5,
         r'$T_{\rm wet}$ $(^\circ {\rm C})$',
         fontsize=15,
         color='blue')
plt.text(-2, 6, r'$\mathcal{R}_{\rm H}$', fontsize=15)
plt.ylim(top=18, bottom=-1)
plt.xlim(right=40)
plt.ylabel(r'$w$ $({\rm g/kg})$')

plt.tight_layout()
switch_mpl_style(latex=False)
```

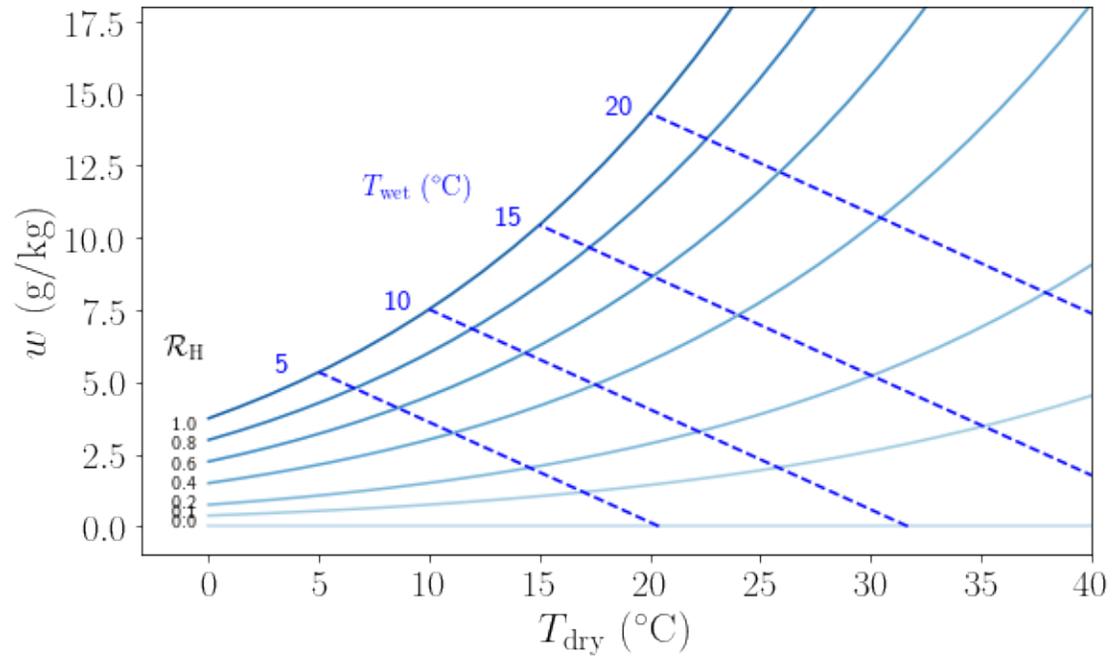



\end{document}